\RequirePackage{fix-cm}
\documentclass[smallextended]{svjour3}       
\smartqed  

\usepackage{hyperref}
\usepackage{graphicx}
\begin{document}

\title{Spectra and decay properties of $\Lambda_b$ and $\Sigma_b$ baryons
}


\author{Amee Kakadiya         \and
        Zalak Shah    \and 
        Keval Gandhi      \and
        Ajay Kumar Rai
}


\institute{Department of Physics, Sardar Vallabhbhai National Institute of Technology, Surat-395007, Gujarat, India. \at
           \and
           Zalak Shah\at
              \email{zalak.physics@gmail.com}
}

\date{Received: date / Accepted: date}

\maketitle

\begin{abstract}
In this article, we study the radial states as well as orbital states mass spectra of the nonstrange baryons from bottom family utilizing hypercentral Constituent Quark Model(hCQM), a non-relativistic approach by employing screened potential as confining potential with color Coulomb potential. The spin-spin, spin-orbital and tensor interaction are considered for hyperfine splitting calculation. The mass spectra calculated and compared with other theoretical approaches. The Regge trajectories has been plotted in $(J, M^2)$ plane using the calculated mass spectra. Further, the single-pion decay and radiative decay have been described and magnetic moments are determined for the ground state.
\end{abstract}

\section{Introduction}
\label{intro}
The study of singly heavy baryon spectrum is quite helpful to extend the understanding of Quantum Chromodynamics(QCD), because the heavy quark inside the baryon behaves as a tool to understand the nature of interaction between the quarks. In  recent  years,  a  huge  number  of  both  experimental  and  theoretical  studies done on properties of heavy baryons and the masses of ground and excited states of charm and bottom baryons containing a single heavy quark \textit{b} or \textit{c} have been published. Many decay channels of the bottom baryons have been observed in the operation of LHCb and numerous data on heavy baryons have been collected. Nonstrange singly bottom baryons consist of one bottom quark and two light quarks($u$ and $d$). In the constituent quark model, bottom baryons form multiplets according to the internal symmetries of spin, flavour and parity \cite{LHCb123}. \\

$\Lambda_b^{0}$ is the lightest bottom baryon having singlet state caused by isospin 0, containing up, down and bottom quarks inside it. And the other combinations of these three quarks form a triplet having the isospin value 1, which are: $uub(\Sigma_b^{+})$, $udb(\Sigma_b^{0})$ and $ddb(\Sigma_b^{-})$. In recent years, several baryon states has been observed experimentally \cite{PDG}. Singly bottom baryon mostly decay by strong interaction. The singly bottom baryon decay into the lowest bottom baryon state $\Lambda_b^0$ dominantly. The two narrow states, $\Lambda_b(5912)^{0}$ and $\Lambda_b(5920)^{0}$, already observed by the decay of excited states into final state $\Lambda_b^{0} \pi^{+} \pi^{-}$ by LHCb and  the heavier state of these confirmed by CDF collaboration \cite{CDF2012} and  later by LHCb \cite{LHCb123}. Mass prediction of ground states and other higher excited states has been studied in many theoretical works and still going on \cite{Moosavi,Azizi102,Liang2020,Yang,Wang2021,Azizi101,XChen,Thakkar2017,Zahra2011,Ebert2011,BChen,CapsticIsgur,Jia,Kim,ShahEPJA2016,
Yamaguchi2015,Zhong,Namekawa,Brown,Alexandrou,Padmanath,Can,Mannel}. Recently, two narrow states $\Lambda_b(6146)^0$ and $\Lambda_b(6152)^0$ have been observed by LHCb collaboration in the decay mode $\Lambda_b^{0} \pi^{+} \pi^{-}$ \cite{LHCb123}. The narrow splitting $\Lambda_b(6146)^0$ and $\Lambda_b(6152)^0$ are declared as doublet of $1D$ states having $J^P$ vlaue $\frac{3}{2}^+$ and $\frac{5}{2}^+$ respectively by LHCb collaboration.

\begin{table}
\caption{Mass, width and $J^P$ value of the non-strange singly bottom baryons from \textbf{PDG \cite{PDG}}}
\label{Tab:1}       
\begin{tabular}{llll}
\hline\noalign{\smallskip}
Resonance & Mass(MeV) & Width(MeV) & $J^P$ \\
\noalign{\smallskip}\hline\noalign{\smallskip}
$\Lambda_{b}^{0}$ & $5619.60\pm 0.17$ & - & $\frac{1}{2}^{+}$\\
$\Lambda_{b}(5912)^{0}$ & $5912.20\pm 0.13\pm 0.17$ & $<0.66$ & $\frac{1}{2}^{-}$ \\
$\Lambda_{b}(5920)^{0}$ & $5919.92\pm 0.19$ & $<0.63$ & $\frac{3}{2}^{-}$ \\
$\Lambda_{b}(6146)^{0}$ & $6146.17\pm 0.33\pm 0.27$ & $2.9\pm 1.3\pm 0.3$ & $\frac{3}{2}^{+}$ \\
$\Lambda_{b}(6152)^{0}$ & $6152.51\pm 0.26\pm 0.27$ & $2.1\pm 0.8\pm 0.3$ & $\frac{5}{2}^{+}$ \\
$\Sigma_{b}^{+}$ & $5810.56\pm 0.25$ & $5.0\pm 0.5$ & $\frac{1}{2}^{+}$ \\
$\Sigma_{b}^{-}$ & $5815.64\pm 0.27$ & $5.3\pm 0.5$ & $\frac{1}{2}^{+}$ \\
$\Sigma_{b}^{*+}$ & $5830.32\pm 0.27$ & $9.4\pm 0.5$ & $\frac{3}{2}^{+}$ \\
$\Sigma_{b}^{*-}$ & $5834.74\pm 0.30$ & $10.4\pm 0.8$ & $\frac{3}{2}^{+}$ \\
$\Sigma_{b}(6097)^{+}$ & $6095.8\pm 1.7\pm 0.4$ & $31.0\pm 5.5\pm 0.7$ & $?^?$ \\
$\Sigma_{b}(6097)^{-}$ & $6098.0\pm 1.7\pm 0.5$ & $28.9\pm 4.2\pm 0.9$ & $?^?$ \\
$\Sigma_{b}(6227)^{-}$ & $6226.9\pm 2.0\pm 0.4$ & $18.1\pm 5.4\pm 1.8$ & $?^?$ \\
\noalign{\smallskip}\hline
\end{tabular}
\end{table}

 $\Sigma_b(6097)^{\pm}$ states observed by LHCb experiment at $\sqrt{s}=$ 7 to 8 TeV corresponding to an integrated luminosity $3 fb^{-1}$ using \textit{pp} collision \cite{LHCb122}.  The $J^P$ value of this state has not been confirmed yet. The identification of $J^P$ value and other properties of the new state make it interesting and challenging.                        

The mass spectra of singly bottom baryons have been studied in different approaches: QCD sumrules \cite{Wang2021,Azizi101,XChen}, hypercentral constituent quark model \cite{Thakkar2017,Zahra2011}, relativistic quark-diquark model \cite{Ebert2011}, relativistic flux tube model \cite{BChen}, relativistic quark potential model\cite{CapsticIsgur},rotating QCD string model \cite{Jia}, non-relativistic two-body potential model with hybrid approach \cite{Kim},  heavy quark limit in the one-boson exchange potential \cite{Yamaguchi2015}, the chiral quark model \cite{Zhong}, lattice QCD \cite{Namekawa,Brown,Alexandrou,Padmanath,Can}, HQET(heavy quark effective theory) \cite{Mannel} etc. In this paper, the masses of radial and orbital states of non-strange singly bottom baryons have been calculated using hypercentral constituent quark model (hCQM), a non-relativistic approach with employing screening potential as confining potential. The mass spectra is presented \textbf{for orbital quantum number $l=0$, $l=1$, $l=2$ and $l=3$ }states. Using this mass spectra, Regge trajectories has been plotted in $(J, M^2)$ plane.

This paper is organized as follows: Hypercentral constituent quark model, the theoretical approach has been presented in detail in Section \ref{sec:section 2}  after the introduction. In Section \ref{sec:section 3}, mass spectra of $\Lambda_b^{0}$ has been listed in Table \ref{Tab:Table 2} and mass spectra of $\Sigma_b^{\pm}$ has been listed in Table \ref{Tab:Table 3}, Table \ref{Tab:Table 4} and Table \ref{Tab:Table 5}. Analysis of mass spectra and  Regge trajectory plots have been presented after the mass spectra. The properties of non-strange singly heavy baryons like, magnetic moment, electromagnetic decay, strong decay and its decay widths has been presented. And the summary of presented work is in the last section.

\section{Theoretical Model}
\label{sec:section 2}
In this section, we explain the theoretical framework which describes the interaction inside the baryonic system with one heavy(bottom) and two light($u$ and $d$) quarks. The hypercentral Constituent Quark Model(hCQM), a non-relativistic approach, has been used to explain how the constituent quarks interact inside the nonstrange singly heavy bottom baryons. Screened potential has been employed as confining potential with color-coulomb potential in spin-independent term, which yields a confining effect of quark separation at extent distance. The relative coordinates availed to describe the interaction inside the three quark baryonic system called Jacobi coordinates($\vec{\rho}$ and $\vec{\lambda}$) \cite{Giannini}.

\begin{equation}
\label{eqn:1} 
\vec{\rho}=\frac{\vec{r_1}-\vec{r_2}}{\sqrt2} \hspace{0.5cm} and \hspace{0.5cm} 
\vec{\lambda}=\frac{\vec{r_1}+\vec{r_2}-2\vec{r_3}} {\sqrt{6}}.
\end{equation} 

\noindent The Hamiltonian of the three quark bound system is given by \cite{ShahEPJA2016},

\begin{equation}
\label{eqn:2} 
H=\frac{P{_x}^2}{2m} + V(x)
\end{equation}

\noindent Here, $x$ is the six-dimensional hypercentral coordinate and $m$ is the reduced mass of the system, which expressed: \begin{equation}
\label{eqn:3}
    m=\frac{2m_\rho m_\lambda}{m_\rho + m_\lambda}
\end{equation}
where, $m_\rho$ and $m_\lambda$ are \cite{Bijkar2000},
\begin{equation}
\label{eqn:4}
m_{\rho}=\frac{2m_1m_2}{m_1+m_2} 
\hspace{0.5cm} and \hspace{0.5cm}
m_{\lambda}=\frac{2m_3(m_{1}^{2}+m_{2}^{2}+m_{1}m_{2})}{(m_1+m_2)(m_1+m_2+m_3)}
\end{equation}

\noindent The constituent quark masses has been considered here to enumerate the masses of these baryons, which are: $m_u=0.338$ GeV, $m_d=0.350$ GeV and $m_b=4.670$ GeV \cite{ShahEPJA2016}.
\noindent The hypercentral coordinate(hyperradias $x$ and hyperangle $\xi$) in terms of Jacobi coordinates can be expressed as below \cite{Giannini}:

\begin{equation}
\label{eqn:5}
x=\sqrt{\rho^2+\lambda^2} 
\hspace{0.5cm} and \hspace{0.5cm}
\xi=arctan\left(\frac{\rho}{\lambda}\right)
\end{equation}



\noindent Now, for the potential term in Hamiltonian, the screened potential has been implemented as confining potential with the color-Coulomb potential. The non-relativistic interaction potential $V(x)$ inside the baryonic system comes in two terms i.e. spin dependent $V_{SD}(x)$ and spin independent $V_{SI}(x)$ potential term. The spin independent potential $V_{SI}(x)$ is summation of Lorentz vector(color-Coulomb) $V_{Col}(x)$ and Lorentz scalar(confining) term $V_{conf}(x)$ terms.

\begin{equation}
\label{eqn:8}
V_{SI}(x)=V_{Col}(x)+V_{conf}(x)
\end{equation}

\noindent where, $V_{Col}(x)$ is the QCD potential interacting between two quarks in three quark baryonic system which is non-relativistic. It can be expressed as 

\begin{equation}
\label{eqn:9}
    V_{Col}(x)=- \frac{\tau}{x}
\end{equation}
 
 and 
 
\begin{equation}
\label{eqn:11}
    V_{conf}(x)=a\left(\frac{1-e^{-{\mu} x}}{\mu}\right)
\end{equation}

\noindent where,$x$ indicates the inter-quark separation, $\frac{2}{3}$ is color factor for baryon and the negative sign with it is indicating the attractive strong interaction between two quarks. And $a$ is the string tension and the constant $\mu$ is the the screening factor. When $x\ll \frac{1}{\mu}$, the screened potential becomes linear like potential $ax$ and when $x\gg \frac{1}{\mu}$, it will be a constant $\frac{a}{\mu}$. Hence, it is intriguing to generate the mass spectra employing screened potential, which gives the lower mass values of excited states than linear potential \cite{Wang2019,Li2009}. The parameter $\alpha_s$ corresponds to the strong running coupling constant.  Screened potential implemented as the confining potential in this work.



\noindent The six-dimensional Schr\"{o}dinger equation defining the quark interaction inside the baryonic system expressed as,

\begin{equation} 
\left[-\frac{1}{2m}\frac{d^2}{dx^2}+\frac{\frac{15}{4}+\gamma(\gamma+4)}{2mx^2}+ V(x)\right]\phi_(x) = E_{B} \phi_{\gamma}(x);
\end{equation}

\noindent where, m is the reduced mass of the baryonic system, $E_B$ is the binding energy of the baryonic states, and $V(x)=V_{SI}(x)+V_{SD}(x)$ is the potential of the baryonic system. The Schr\"{o}dinger equation has been solved using Mathematica notebook \cite{Lucha1999}. The spin average masses are ascertained by taking a summation of constituent quark masses and its binding energy \cite{Gandhinon-strange}.The masses of radial and orbital excited states of three quark baryonic system has been calculated this way and the analysis part of the predicted masses is described in next section.






\section{Mass Spectra and Regge Trajectories}
\label{sec:section 3}
The masses of radial and orbital states of non-strange singly bottom baryons has been calculated by using non-relativistic framework i.e. hypercentral constituent quark model. The screened potential has been used as confining potential with color-Coulomb potential. Table \ref{Tab:Table 2}, Table \ref{Tab:Table 3}, Table \ref{Tab:Table 4} and Table \ref{Tab:Table 5} present the calculated masses and compared with the data of other theoretical predictions. Regge trajectories can be utilized for assigning the possible quantum number of the further hadronic states which can be identify the higher excited states of the heavy baryons.

\begin{table}
\centering
\caption{Predicted masses of $\Lambda_{b}^{0}$ baryon (in GeV)}
\label{Tab:Table 2}       
\begin{tabular}{lllllllllllllll}
\hline\noalign{\smallskip}
State& $J^P$ & Present & PDG \cite{PDG} & \cite{Ebert2011} & \cite{Roberts} & \cite{WHLiang} & \cite{Jia} & \cite{Kim} & \cite{Moosavi} & \cite{Azizi101} & \cite{Liang2020} & \cite{Wang2021} & \cite{Azizi102} \\
\noalign{\smallskip}\hline\noalign{\smallskip}

1S & $\frac{1}{2}^+$ & 5.620 & 5.619 & 5.620 & 5.612 & & 5.615 & 5.620 & 5.620 & & & 5.61 & 5.611\\
2S & $\frac{1}{2}^+$ & 6.067 & & 6.089 & 6.107 & & & & 6.078 & & 6.045 &  6.08 & 6.073\\
3S & $\frac{1}{2}^+$ & 6.386 & & 6.455 & &&&& 6.449 \\
4S & $\frac{1}{2}^+$ & 6.685 & & 6.756 & &&&& 6.752 \\
5S & $\frac{1}{2}^+$ & 6.968 & &7.015 & &&&& 7.023\\
\\
 
$1P$ & $\frac{1}{2}^-$ & 6.056 & 5.912 & 5.930 & 5.939 & 5.820 & 5.908 & 6.079 & 5.919 & & 6.100 & & 5.911\\
$1P$ & $\frac{3}{2}^-$ & 5.952 & 5.919  & 5.942 & 5.941 & 5.969 & 5.921 & 5.923 & 5.929 & & 6.185 \\
\\
$2P$ & $\frac{1}{2}^-$ & 6.218 & & 6.326 & 6.180 &&&& 6.331\\
$2P$ & $\frac{3}{2}^-$ & 6.204 & & 6.333 & 6.191 &&&& 6.325\\
\\
$3P$ & $\frac{1}{2}^-$ & 6.478 & & 6.645 & &&&& 6.641\\
$3P$ & $\frac{3}{2}^-$ & 6.472 & & 6.651 & &&&& 6.647 \\
\\
$4P$ & $\frac{1}{2}^-$ & 6.738 & & 6.917 & &&&& 6.964 \\
$4P$ & $\frac{3}{2}^-$ & 6.736 & & 6.922 & &&&& 6.930 \\
\\
$5P$ & $\frac{1}{2}^-$ & 6.995 & & 7.157 & &&&& 7.163\\
$5P$ & $\frac{3}{2}^-$ & 6.994 & & 7.171 & &&&& 7.182\\

\\
$1D$ & $\frac{3}{2}^+$ & 6.121 & 6.146 & 6.190 & 6.181& & 6.144 & &6.199 & 6.144 \\
$1D$ & $\frac{5}{2}^+$ & 6.119 & 6.152 & 6.196 & 6.183 & &6.152 && 6.210\\
\\
$2D$ & $\frac{3}{2}^+$ & 6.400 & & 6.526 & 6.401 &&&& 6.531\\
$2D$ & $\frac{5}{2}^+$ & 6.399 & & 6.531 & 6.422 &&&& 6.538\\
\\
$3D$ & $\frac{3}{2}^+$ & 6.673 & & 6.811 & &&&& 6.817\\
$3D$ & $\frac{5}{2}^+$ & 6.672 & & 6.814 & &&&& 6.822\\
\\
$4D$ & $\frac{3}{2}^+$ & 6.940 & & 7.060 & &&&& 7.069\\
$4D$ & $\frac{5}{2}^+$ & 6.939 & & 7.063 & &&&& 7.071\\
\\

$1F$ & $\frac{5}{2}^-$ & 6.303 & & 6.408 & 6.206 & & && 6.421 && 6.205 \\
$1F$ & $\frac{7}{2}^-$ & 6.299 & & 6.411 & &&&& 6.423 \\
\\
$2F$ & $\frac{5}{2}^-$ & 6.589 & & 6.705 & &&&& 6.719\\
$2F$ & $\frac{7}{2}^-$ & 6.587 & & 6.708 & &&&& 6.715\\

\noalign{\smallskip}\hline
\end{tabular}
\end{table}

\noindent The Regge trajectories are plotted using the  data presented in Table \ref{Tab:Table 2}, Table \ref{Tab:Table 3}, Table \ref{Tab:Table 4} and Table \ref{Tab:Table 5} in $(J, M^2)$ plane using the equation,
\begin{equation}
    J=\alpha M^2+\alpha_0
\end{equation}
where, $\alpha$ and $\alpha_0$ are the slope of linear fitting line and interception on the $y$-axis respectively. The square of calculated masses for particular $J^P$ values are indicated by various symbols in $(J, M^2)$ plane(see Fig. \ref{Fig:Figure 1}, Fig. \ref{Fig:Figure 2}, Fig. \ref{Fig:Figure 3}) for particular baryonic system.

\subsection{$\Lambda_b^{0}$ State}
$\Lambda_b^{0}$ is the lowest state of bottom baryon family, containing $u$, $d$, and $b$ quarks as constituents with isospin value 0, which causes only siglet state of $\Lambda_b^{0}$ baryon. LHCb collaboration \cite{LHCb123} measured $\Lambda_b^{0}$ state by various decay channels of $\Lambda_b^{0}$ during $pp$ collision at CERN. CDF collaboration \cite{CDF2007} also measured the same state while observing $J/\Psi$ $\Lambda$ reconstructed decay channel of $\Lambda_b^{0}$ during $p \bar{p}$ collision with energy of $1.96$ TeV. Later, $\Lambda_b^{0}$ state confirmed by other experiments such as DELPHI collaboration\ cite{DELPHI} and ALEPH collaboration\ cite{ALEPH}.
The ground state($1S$) mass of $\Lambda_b^{0}$ has been fixed as per experimentally observed mass value 5.620 GeV \cite{PDG} and then calculated the whole mass spectra for other excited states using Mathematica notebook program \cite{Lucha1999}. Presented mass for $J^P=\frac{3}{2}^-$ of $1P$ state 5.952 GeV is very close to the experimental result 5.920 GeV by PDG \cite{PDG} with difference of only 32 MeV and also in good agreement of references  \cite{Thakkar2017} and \cite{Ebert2011}. Our predicted masses of higher $p$-wave states are also compatible with masses calculated by relativistic approach \cite{Ebert2011} and hyper Coulomb plus linear potential \cite{Thakkar2017}.

\noindent In 2019, LHCb \cite{LHCb123} observed the states with $J^P$ value $\frac{3}{2}^+$ and $\frac{5}{2}^+$ are also compatible with our prediction with negligible mass difference of 25 MeV and 33 MeV respectively. The higher $d$-wave state masses are near to the data copared in Table \ref{Tab:Table 2}. The values are lesser than relativistic approach \cite{Ebert2011} masses which indicates the screening effect, the effect of
the confining potential used in model.

\begin{figure}
\centering
\includegraphics[scale=0.8]{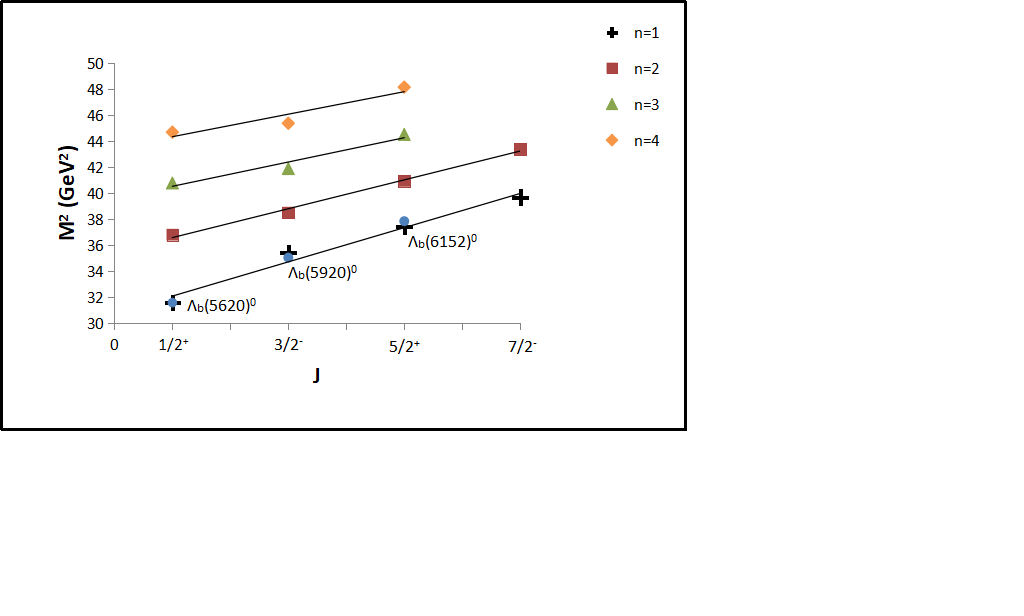}
\caption{The $M^2 \rightarrow J$ Regge trajectory of $\Lambda_{b}^0$ baryon with natural parity.}
\label{Fig:Figure 1}       
\end{figure}

\noindent Fig. \ref{Fig:Figure 1} shows the Regge trajectory lines of $\Lambda_b^{0}$. The black plus sign denotes the $M^2$ values for $n=1$, red square denotes the $M^2$ values for $n=2$, green trangle denotes the $M^2$ values for $n=3$, yellow square denotes the $M^2$ values for $n=4$ and blue dots denote the experimentally observed states. Regge lines are parallel at first sight, but the lines will intersect if they would be extended at very long distance, which is effect of screened potential at higher excited states.

\subsection{$\Sigma_b^{\pm}$ States}
Four of six $\Sigma_b$ states, $\Sigma_b^{\pm}$ and $\Sigma_b^{*\pm}$ states are observed by CDF-II detector \cite{CDF2012,CDF2007} in Fermi Lab and later reported briefly by LHCb \cite{LHCb122}. In present study, masses of $\Sigma_b^{\pm}$ and $\Sigma_b^{*\pm}$ states has been calculated by considering $u$ and $d$ quarks have different masses. Here, the ground state($1S$) of $\Sigma_b^{+}$ and $\Sigma_b^{-}$ have been fixed as per the experimentally observed mass value given by PDG \cite{PDG} and the further states masses enumerated with Mathematica notebook program \cite{Lucha1999}. $\Sigma_b$ isotriplet has two spin states:$\frac{1}{2}$ and $\frac{3}{2}$.The masses of radial states of $J^P$ value $\frac{1}{2}$ and $\frac{3}{2}$ have been listed in Table \ref{Tab:Table 3}. The predicted masses of further radial states are in accordance with the references \cite{Ebert2011}, \cite{Jia}, \cite{Roberts} and \cite{Bhavsar}.
\noindent Recently, the resonance states $\Sigma_b(6097)^{+}$ and $\Sigma_b(6097)^{-}$ has been found by LHCb \cite{PDG}, which have no confirmed $J^P$ value. As per our calculation, it may be $1P(\frac{5}{2}^{-})$ of $\Sigma_b^{*+}$ or $1P(\frac{1}{2}^{-})$ of $\Sigma_b^{*-}$ (shown bold numbers in Table \ref{Tab:Table 4} and  Table \ref{Tab:Table 5}). For $d$ and $f$ states, predicted masses are quite near and lesser than the reference \cite{Ebert2011}. For higher values of $n$(principal quantum number), the difference increases which is caused by the confining potential(screened potential).

\begin{figure}
\centering
\includegraphics[scale=0.8]{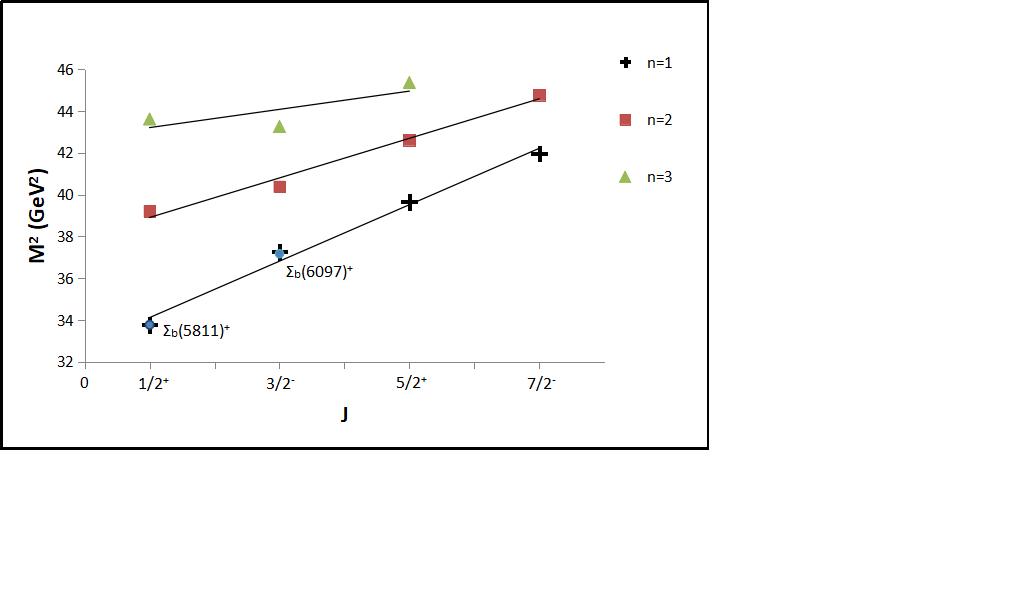}
\caption{The $M^2 \rightarrow J$ Regge trajectory of $\Sigma_{b}^+$ baryon with natural parity.}
\label{Fig:Figure 2}
\end{figure}

\begin{figure}
\centering
\includegraphics[scale=0.8]{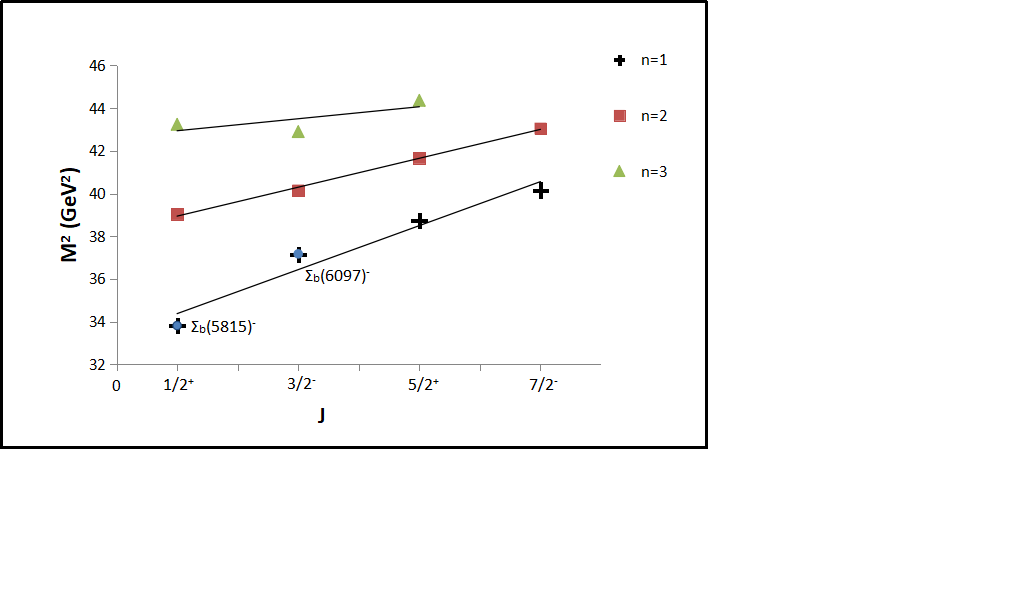}
\caption{The $M^2 \rightarrow J$ Regge trajectory of $\Sigma_{b}^-$ baryon with natural parity.}
\label{Fig:Figure 3}
\end{figure}

Fig. \ref{Fig:Figure 2} and Fig. \ref{Fig:Figure 3} are the Regge trajectories of $\Sigma_b^{+}$ and $\Sigma_b^{-}$ respectively. The black plus sign denotes the $M^2$ values for $n=1$, red square denotes the $M^2$ values for $n=2$, green trangle denotes the $M^2$ values for $n=3$, and blue dots denote the experimentally observed states. Regge lines are about to intersect in both plot, which means effect of screening potential increasing for higher states.

\begin{table}
\centering
\caption{Predicted masses of the radial excited states of $\Sigma_{b}^{\pm}$ baryons (in GeV)}
\label{Tab:Table 3}       
\begin{tabular}{llllllllll}
\hline\noalign{\smallskip}
Particle & state & $J^P$ & Present & PDG \cite{PDG} & \cite{Ebert2011} & \cite{Roberts} & \cite{Jia} & \cite{Bhavsar} & \cite{Moosavi} \\
\noalign{\smallskip}\hline\noalign{\smallskip}
$\Sigma_{b}^+$ & $1S$ & $\frac{1}{2}^+$ & 5.811 & 5.81056 & 5.808 & 5.833 & 5.825 &5.815 \\  
           & $2S$ & $\frac{1}{2}^+$ & 6.262 &  & 6.213 & 6.294 &&& 6.218\\
           & $3S$ & $\frac{1}{2}^+$ & 6.605 &  & 6.575 &&&& 6.579\\ 
           & $4S$ & $\frac{1}{2}^+$ & 6.927 &  & 6.869 &&&& 6.863\\  
           & $5S$ & $\frac{1}{2}^+$ & 7.231 &  & 7.124 &&&& 7.128\\ 
\\ 
& $1S$ & $\frac{3}{2}^+$ & 5.832 & 5.83032 & 5.834 & 5.858 & & 5.828\\ 
& $2S$ & $\frac{3}{2}^+$ & 6.278 &  & 6.226 & 6.308 &&& 6.232\\  
& $3S$ & $\frac{3}{2}^+$ & 6.614 &  & 6.583 &&&& 6.589 \\  
& $4S$ & $\frac{3}{2}^+$ & 6.933 &  & 6.876 &&&& 6.881\\  
& $5S$ & $\frac{3}{2}^+$ & 7.235 &  & 7.129 &&&& 7.130\\ 

\\\hline
\\
$\Sigma_{b}^-$ & $1S$ & $\frac{1}{2}^+$ & 5.815 & 5.81564 & 5.808 & 5.833 & 5.825 & 5.740\\
           & $2S$ & $\frac{1}{2}^+$ & 6.248 &  & 6.213 & 6.294\\
           & $3S$ & $\frac{1}{2}^+$ & 6.576 & & 6.575\\
           & $4S$ & $\frac{1}{2}^+$ & 6.885 & & 6.869\\
           & $5S$ & $\frac{1}{2}^+$ & 7.176 & & 7.124\\
\\
& $1S$ & $\frac{3}{2}^+$ & 5.832 & 5.83474 & 5.834 & 5.858 & & 5.829\\ 
& $2S$ & $\frac{3}{2}^+$ & 6.268 &  & 6.226 & 6.308\\  
& $3S$ & $\frac{3}{2}^+$ & 6.596 &  & 6.583 \\  
& $4S$ & $\frac{3}{2}^+$ & 6.907 &  & 6.876 \\  
& $5S$ & $\frac{3}{2}^+$ & 7.203 &  & 7.129 \\

\noalign{\smallskip}\hline
\end{tabular}
\end{table}

\begin{table}
\centering
\caption{Predicted masses of the orbital excited states of $\Sigma_{b}^{+}$ baryons (in GeV)}
\label{Tab:Table 4}       
\begin{tabular}{llllllllll}
\hline\noalign{\smallskip}
state & $J^P$ & Present & \cite{Ebert2011} & \cite{Roberts} & \cite{WHLiang} &\cite{Moosavi} & \cite{Liang2020}& \cite{Yang} & \cite{Jia} \\
\noalign{\smallskip}\hline\noalign{\smallskip}
 $1P$ & $\frac{1}{2}^-$ & 6.104 & 6.101 & 6.099 & 6.179& 6.122& 6.070 & $6.06\pm 0.13$ & 6.088 \\  
$1P$ & $\frac{3}{2}^-$ & 6.100 & 6.096 & 6.101& & 6.091 & 6.070 & $6.07\pm 0.13$\\  
$1P$ & $\frac{1}{2}^-$ & 6.106 & 6.095 & & 6.063 && 6.070\\  
$1P$ & $\frac{3}{2}^-$ & 6.102 & 6.087 &&& 6.085 && $6.11\pm 0.16$\\  
$1P$ & $\frac{5}{2}^-$ & \textbf{6.097} & 6.084 &&& 6.090& &$6.12\pm0.15$\\ 
\\
$2P$ & $\frac{1}{2}^-$ & 6.355 & 6.440 & 6.106 & 6.477 & 6.453\\  
$2P$ & $\frac{3}{2}^-$ & 6.353 & 6.430 & 6.105 & & 6.442\\  
$2P$ & $\frac{1}{2}^-$ & 6.356 & 6.430 & & 6.491\\  
$2P$ & $\frac{3}{2}^-$ & 6.354 & 6.423 \\  
$2P$ & $\frac{5}{2}^-$ & 6.351 & 6.421 \\ 
\\
$3P$ & $\frac{1}{2}^-$ & 6.578 & 6.756 &&& 6.760\\  
$3P$ & $\frac{3}{2}^-$ & 6.577 & 6.742 &&& 6.747\\  
$3P$ & $\frac{1}{2}^-$ & 6.579 & 6.096 \\  
$3P$ & $\frac{3}{2}^-$ & 6.577 & 6.736 \\ 
$3P$ & $\frac{5}{2}^-$ & 6.575 & 6.732 \\ 
\\
$4P$ & $\frac{1}{2}^-$ & 6.778 & 7.024 &&& 7.028\\  
$4P$ & $\frac{3}{2}^-$ & 6.777 & 7.009 &&& 7.013\\ 
$4P$ & $\frac{1}{2}^-$ & 6.779 & 7.008 \\
$4P$ & $\frac{3}{2}^-$ & 6.778 & 7.003 \\ 
$4P$ & $\frac{5}{2}^-$ & 6.776 & 6.999 \\
\\
$1D$ & $\frac{3}{2}^+$ & 6.298 & 6.326 &&& 6.329\\
$1D$ & $\frac{5}{2}^+$ & 6.294 & 6.284 & 6.325 && 6.288\\
$1D$ & $\frac{1}{2}^+$ & 6.303 & 6.311 \\
$1D$ & $\frac{3}{2}^+$ & 6.300 & 6.285 \\
$1D$ & $\frac{5}{2}^+$ & 6.295 & 6.270 \\
$1D$ & $\frac{7}{2}^+$ & 6.290 & 6.260 & 6.333 \\
\\
$2D$ & $\frac{3}{2}^+$ & 6.529 & 6.647 &&&6.650 \\
$2D$ & $\frac{5}{2}^+$ & 6.526 & 6.612 & 6.328 && 6.601\\
$2D$ & $\frac{1}{2}^+$ & 6.533 & 6.636 \\
$2D$ & $\frac{3}{2}^+$ & 6.530 & 6.612 \\
$2D$ & $\frac{5}{2}^+$ & 6.527 & 6.598 \\
$2D$ & $\frac{7}{2}^+$ & 6.524 & 6.590 & 6.554\\
\\

$3D$ & $\frac{3}{2}^+$ & 6.736 & \\
$3D$ & $\frac{5}{2}^+$ & 6.734 & \\
$3D$ & $\frac{1}{2}^+$ & 6.738 & \\
$3D$ & $\frac{3}{2}^+$ & 6.736 & \\
$3D$ & $\frac{5}{2}^+$ & 6.735 & \\
$3D$ & $\frac{7}{2}^+$ & 6.733 & \\
\\
$4D$ & $\frac{3}{2}^+$ & 6.922 & \\
$4D$ & $\frac{5}{2}^+$ & 6.921 & \\
$4D$ & $\frac{1}{2}^+$ & 6.923 & \\
$4D$ & $\frac{3}{2}^+$ & 5.922 & \\
$4D$ & $\frac{5}{2}^+$ & 6.921 & \\
$4D$ & $\frac{7}{2}^+$ & 6.920 & \\
\\
\hline
\end{tabular}\\
{continued...}
\end{table}

\begin{table}
\centering
\addtocounter{table}{-1}
\begin{tabular}{llllll}
\hline\noalign{\smallskip}
state & $J^P$ & Present & \cite{Ebert2011} &\cite{Roberts} &
\cite{Moosavi}\\
\noalign{\smallskip}\hline\noalign{\smallskip}

$1F$ & $\frac{5}{2}^-$ & 6.479 & 6.564 & 6.172 \\
$1F$ & $\frac{7}{2}^-$ & 6.474 & 6.500 \\
$1F$ & $\frac{3}{2}^-$ & 6.485 & 6.550 && 6.556\\
$1F$ & $\frac{5}{2}^-$ & 6.481 & 6.501 && 6.569\\
$1F$ & $\frac{7}{2}^-$ & 6.476 & 6.472 && 6.508\\
$1F$ & $\frac{9}{2}^-$ & 6.469 & 6.459 \\
\\
$2F$ & $\frac{5}{2}^-$ & 6.692 \\
$2F$ & $\frac{7}{2}^-$ & 6.689 \\
$2F$ & $\frac{3}{2}^-$ & 6.695 \\
$2F$ & $\frac{5}{2}^-$ & 6.693 \\
$2F$ & $\frac{7}{2}^-$ & 6.690 \\
$2F$ & $\frac{9}{2}^-$ & 6.687 \\
\noalign{\smallskip}\hline
\end{tabular}
\end{table}

\begin{table}
\centering
\caption{Predicted masses of the orbital excited states of $\Sigma_{b}^{-}$ baryons (in GeV)}
\label{Tab:Table 5}       
\begin{tabular}{llllllllll}
\hline\noalign{\smallskip}
state & $J^P$ & Present & \cite{Ebert2011} & \cite{Roberts} & \cite{WHLiang} & \cite{Liang2020}& \cite{Yang} & \cite{Jia}\\
\noalign{\smallskip}\hline\noalign{\smallskip}
$1P$ & $\frac{1}{2}^-$ & 6.095 & 6.101 & 6.099 & 6.179 & 6.070& $6.06\pm0.13$ & 6.088\\  
$1P$ & $\frac{3}{2}^-$ & 6.092 & 6.096 & 6.101 && 6.070 &$6.07\pm0.13$\\  
$1P$ & $\frac{1}{2}^-$ & \textbf{6.097} & 6.095 & & 6.063 & 6.070  \\  
$1P$ & $\frac{3}{2}^-$ & 6.094 & 6.087 &&& 6.085 &$6.11\pm0.16$ \\ 
$1P$ & $\frac{5}{2}^-$ & 6.090 & 6.084 &&& 6.090 &$6.12\pm0.15$\\ 
\\
$2P$ & $\frac{1}{2}^-$ & 6.336  & 6.440 & 6.106 & 6.477\\  
$2P$ & $\frac{3}{2}^-$ & 6.334  & 6.430 & 6.105 \\  
$2P$ & $\frac{1}{2}^-$ & 6.337  & 6.430 & & 6.491\\
$2P$ & $\frac{3}{2}^-$ & 6.335  & 6.423 \\  
$2P$ & $\frac{5}{2}^-$ & 6.333  & 6.421 \\ 
\\
$3P$ & $\frac{1}{2}^-$ & 6.550 & 6.756 \\  
$3P$ & $\frac{3}{2}^-$ & 6.549 & 6.742 \\  
$3P$ & $\frac{1}{2}^-$ & 6.551 & 6.096 \\  
$3P$ & $\frac{3}{2}^-$ & 6.550 & 6.736 \\ 
$3P$ & $\frac{5}{2}^-$ & 6.548 & 6.732 \\ 
\\
$4P$ & $\frac{1}{2}^-$ & 6.742 & 7.024 \\  
$4P$ & $\frac{3}{2}^-$ & 6.741 & 7.009 \\ 
$4P$ & $\frac{1}{2}^-$ & 6.742 & 7.008 \\
$4P$ & $\frac{3}{2}^-$ & 6.741 & 7.003 \\ 
$4P$ & $\frac{5}{2}^-$ & 6.740 & 6.999 \\
\\
$1D$ & $\frac{3}{2}^+$ & 6.224 & 6.326 \\
$1D$ & $\frac{5}{2}^+$ & 6.220 & 6.284 & 6.325\\
$1D$ & $\frac{1}{2}^+$ & 6.228 & 6.311 \\
$1D$ & $\frac{3}{2}^+$ & 6.225 & 6.285 \\
$1D$ & $\frac{5}{2}^+$ & 6.221 & 6.270 \\
$1D$ & $\frac{7}{2}^+$ & 6.216 & 6.260 & 6.333\\
\\
$2D$ & $\frac{3}{2}^+$ & 6.455 & 6.647 \\
$2D$ & $\frac{5}{2}^+$ & 6.452 & 6.612 & 6.328\\
$2D$ & $\frac{1}{2}^+$ & 6.458 & 6.636 \\
$2D$ & $\frac{3}{2}^+$ & 6.456 & 6.612 \\
$2D$ & $\frac{5}{2}^+$ & 6.453  & 6.598 \\
$2D$ & $\frac{7}{2}^+$ & 6.450 & 6.590 & 6.554\\
\\
$3D$ & $\frac{3}{2}^+$ & 6.661 & & & & \\
$3D$ & $\frac{5}{2}^+$ & 6.659 & & & & \\
$3D$ & $\frac{1}{2}^+$ & 6.662 & & & & \\
$3D$ & $\frac{3}{2}^+$ & 6.661 & & & & \\
$3D$ & $\frac{5}{2}^+$ & 6.660 & & & & \\
$3D$ & $\frac{7}{2}^+$ & 6.658 & & & & \\
\\
$4D$ & $\frac{3}{2}^+$ & 6.846 & & & & \\
$4D$ & $\frac{5}{2}^+$ & 6.844 & & & & \\
$4D$ & $\frac{1}{2}^+$ & 6.847 & & & & \\
$4D$ & $\frac{3}{2}^+$ & 6.846 & & & & \\
$4D$ & $\frac{5}{2}^+$ & 6.845 & & & & \\
$4D$ & $\frac{7}{2}^+$ & 6.843 & & & & \\
\\
\noalign{\smallskip}\hline\noalign{\smallskip}
\end{tabular}\\
{continued...}
\end{table}

\begin{table}
\centering
\addtocounter{table}{-1}
\begin{tabular}{llllll}
\hline\noalign{\smallskip}
state & $J^P$ & Present & \cite{Ebert2011} &\cite{Roberts} \\
\noalign{\smallskip}\hline\noalign{\smallskip}

$1F$ & $\frac{5}{2}^-$ & 6.337 & 6.564 & 6.172\\
$1F$ & $\frac{7}{2}^-$ & 6.331 & 6.500 \\
$1F$ & $\frac{3}{2}^-$ & 6.343 & 6.550 \\
$1F$ & $\frac{5}{2}^-$ & 6.339 & 6.501 \\
$1F$ & $\frac{7}{2}^-$ & 6.333 & 6.472 \\
$1F$ & $\frac{9}{2}^-$ & 6.326 & 6.459 \\
\\
$2F$ & $\frac{5}{2}^-$ & 6.561 \\
$2F$ & $\frac{7}{2}^-$ & 6.557 \\
$2F$ & $\frac{3}{2}^-$ & 6.566 \\
$2F$ & $\frac{5}{2}^-$ & 6.562 \\
$2F$ & $\frac{7}{2}^-$ & 6.558 \\
$2F$ & $\frac{9}{2}^-$ & 6.554 \\

\noalign{\smallskip}\hline
\end{tabular}
\end{table}

\section{Properties}
\label{sec:4}
\subsection{Magnetic Moment and Radiative Decay}
\noindent 

Numerical values of magnetic moment can be enumerated using the values of spin, charge and effective masses of constituent quarks inside the baryons. An exchange of photons among the non-strange singly bottom baryonic states is called radiative decay, which do not have phase space restriction. Hence, some of such decay modes contribute in the total decay rate significantly. And when the baryonic states decays into any hadronic state is called strong decay, which obeys all conservation laws.

\subsubsection{Magnetic Moment}
The magnetic moment of the baryon is the fundamental and intrinsic property caused of its mass and spin of its constituents. The magnetic moment expression of baryon can be derive by operating the expectation value equation given below \cite{GandhiDecay},

\begin{equation}
    \mu_B=\sum_q \langle \Phi_{sf}|\hat{\mu}_{qz}|\Phi_{sf} \rangle;  \quad \quad q=u, d, b
\end{equation}
Here, $\Phi_{sf}$ represents the spin-flavour wave-function of the baryon and $\hat{\mu}_{qz}$ is the magnetic moment operator. 
\noindent The magnetic moment of individual quark is given by \cite{GandhiDecay},
\begin{equation}
    \mu_q=\frac{e_q}{2m_q^{eff}}\cdot \sigma_q
\end{equation}

where, $e_q$ and $\sigma_q$ are charge and spin of the individual constituent quark of the baryonic system respectively and $m_q^{eff}$ is the effective mass of constituent quark which is expressed as given below \cite{GandhiDecay}:
\begin{equation}
    m_q^{eff}=m_q \left(1+\frac{\langle H \rangle}{\sum_q m_q}\right)
\end{equation}
Here, the Hamiltonian $\langle H \rangle$ is expressed as the difference of predicted mass(in experiment, measured mass) and total of the individual constituent masses of the baryon ($\langle H \rangle=M-\sum_q m_q$; where, $M$ is predicted mass of the particular baryonic state). And $m_q^{eff}$ is the mass of bounded quark inside the baryon with consideration of the interaction with other two quarks. 
The magnetic moment of ground states of all baryonic systems mentioned in Section \ref{sec:section 3} has been calculated by using above equations which is shown in the Table \ref{Tab:Table 6}. In this work, the predicted masses of particular baryonic states shown in Section \ref{sec:section 3} has been used to calculate the magnetic moment for the same. 

\begin{table}
\centering
\caption{Magnetic moments of the non-strange singly bottom baryons}
\label{Tab:Table 6}       
\begin{tabular}{lllll}
\hline\noalign{\smallskip}
Baryon & $J^P$ & Expression & Present & \cite{Majethiya2016} \\
\hline
$\Lambda_{b}^0$ & $\frac{1}{2}^+$ & $\mu_b$ & -0.064 & -0.063\\
$\Sigma_{b}^+$ & $\frac{1}{2}^+$ & $\frac{4}{3}\mu_u-\frac{1}{3}\mu_b$ & 2.295 & 2.377\\
$\Sigma_{b}^-$ & $\frac{1}{2}^+$ & $\frac{4}{3}\mu_d-\frac{1}{3}\mu_b$ & -1.077 & -1.158\\
$\Sigma_{b}^{*+}$ & $\frac{3}{2}^+$ & $2\mu_u+\mu_b$ & 3.338 & 3.461\\
$\Sigma_{b}^{*-}$ & $\frac{3}{2}^+$ & $2\mu_d+\mu_b$ & -1.703 & -1.822 \\
\hline
\end{tabular}
\end{table}

\subsubsection{Radiative Decay}
The electromagnetic radiative decay width is expressed as \cite{GandhiDecay},
\begin{equation}
    \Gamma=\frac{k^3}{4\pi}\frac{2}{2J+1}\frac{e^2}{2m_p^2}\mu_{B\rightarrow B'}^2
\end{equation}
where, $k$ is photon energy, $J$ is total angular momentum of the initial baryonic state, $m_p$ is the mass of proton (in MeV) and $\mu_b$ transition magnetic moment for the particular radiative decay.

The radiative transition magnetic moment can be calculated by the sandwiching spin-flavour wave functions of initial baryon state($B$) and final baryon state($B'$) with $z$ component of magnetic moment operator, which is expressed as below:
\begin{equation}
\mu_{B\rightarrow B'}=\langle \Phi_B|\mu_{B\rightarrow B'}|\Phi_{B'}\rangle    
\end{equation}
The spin-flavour wave function of initial baryon($\Phi_B$) state and final baryon state($\Phi_B'$) can be determined as described in  \cite{Majethiya2009}.
 
The radiative transition decay widths are shown in Table \ref{Tab:Table 7}, which is calculated by using transition magnetic moment for the particular transition.
  
\begin{table}
\centering
\caption{Transition magnetic moment and Decay widths of radiative transitions (in keV)}
\label{Tab:Table 7}       
\begin{tabular}{lllll}
\hline\noalign{\smallskip}
Transition & Transition magnetic moment & Present Decay width & \cite{Majethiya2016} & \cite{Aliev}\\
\hline
$\Sigma_b^{*+} \rightarrow \Sigma_b^+ \gamma$ & 1.664 & 0.13 & 0.11 & 0.46 \\
$\Sigma_b^{*-} \rightarrow \Sigma_b^- \gamma$ & -0.717 & 0.013 & 0.02 & 0.11\\
\hline
\end{tabular}
\end{table}

\subsection{Strong Decay}
Here, $P$-wave, $S$-wave and $D$-wave transition decays has been calculated using the Lagrangian given in ref.\cite{Pirjol}; where, $P$-wave transition is transition between $s$-wave baryons, $S$-wave transition is transition between $s$-wave to $p$-wave($J^P=\frac{1}{2}$) baryons and $D$-wave transition is  transition between $p$-wave($J^P=\frac{3}{2}$,$\frac{5}{2}$) to $s$-wave baryons. The equations of decay rate for each transition are given further in this section(equation \ref{eqn:19}-\ref{eqn:24}) with an example. 
The pion momentum for two body decay $x \rightarrow y + \pi$ is as given below:
\begin{equation}
 p_\pi =\frac{1}{2m_x}\sqrt{[m_{x}^2-(m_y+m_\pi)^2][m_{x}^2-(m_y-m_\pi)^2]}   
\end{equation}

\subsubsection{$P$-wave Transition Decay Rate}
The strong decay rate for the $P$-wave transition $\Sigma_{b}^{+}(1^2S_{\frac{1}{2}})$ to ground state of $\Lambda_{b}^0$ with an exchange of single pion \cite{GandhiDecay},
where, strong coupling constant $a_1=0.612$ \cite{GandhiDecay}, the pion decay constant $f_{\pi}=132$ MeV and $p_{\pi}^3$ represents the $P$-wave transition momentum \cite{GandhiDecay}. Masses of $\Sigma_{b}^{+}$ and $\Lambda_{b}^{0}$ has been taken as shown in Section  \ref{sec:section 3}.

\subsubsection{$S$-wave Transition Decay Rate}
The equation of decay rate for $S$-wave transition into $\Sigma_{b}^{+}(1^2S_{\frac{1}{2}})$ from $\Lambda_{b}^{0}(1^2P_{\frac{1}{2}})$ with a single pion exachange, 

\begin{equation}
\label{eqn:20}
    \Gamma_{\Lambda_{b}^{0}(1^2P_{\frac{1}{2}}) \rightarrow \Sigma_{b}^{+} \pi^{-}}=\frac{b_1^2}{2\pi f_{\pi}^2} \frac{M_{\Sigma_{b}^{+}}}{M_{\Lambda_{b}^{0}(1^2P_{\frac{1}{2}})}} E_{\pi}^2 p_{\pi}
    \end{equation}
The coupling constant $b_1=0.572$\cite{GandhiDecay}, $p_{\pi}$ presents $S$-wave transition momentum and $E_{\pi} \approx m_{\pi}$ for the single pion at rest. The decay rate for $\Sigma_{b}^{+}(1^2P_{\frac{1}{2}}) \rightarrow \Lambda_{b}^0 \pi^+$(coupling constant $b_2=\sqrt{3} \cdot b_1$ \cite{GandhiDecay}) is,
\begin{equation}
\label{eqn:21}
    \Gamma_{\Sigma_{b}^{+}(1^2P_{\frac{1}{2}}) \rightarrow \Lambda_{b}^0 \pi^+}=\frac{b_2^2}{2\pi f_{\pi}^2} \frac{M_{\Lambda_{b}^{0}}}{M_{\Sigma_{b}^{+}(1^2P_{\frac{1}{2}})}} E_{\pi}^2 p_{\pi}
\end{equation}

\subsubsection{$D$-wave Transition Decay Rate}
The decay rate for strong decay into $\Sigma_{b}^{+}$ from $\Lambda_{b}^{0}(1^2P_{\frac{3}{2}})$,
\begin{equation}
\label{eqn:22}
    \Gamma_{\Lambda_{b}^{0}(1^2P_{\frac{3}{2}}) \rightarrow \Sigma_{b}^{+} \pi^{-}}=\frac{2b_3^2}{9\pi f_{\pi}^2} \frac{M_{\Sigma_{b}^{+}}}{M_{\Lambda_{b}^{0}(1^2P_{\frac{3}{2}})}} p_{\pi}^5
    \end{equation}
here, coupling constant $b_3=3.50 \times 10^{-3} MeV^{-1}$  \cite{GandhiDecay} and $p_{\pi}^5$ presents $D$-wave transition momentum.
The decay rate for $D$-wave transition $\Sigma_{b}^{+}(1^2P_{\frac{3}{2}})$(or $J^P=\frac{5}{2}$) to $\Lambda_{b}^0 \pi^+$ is,
\begin{equation}
\label{eqn:23}
    \Gamma_{\Sigma_{b}^{+}(1^2P_{\frac{1}{2}}) \rightarrow \Lambda_{b}^0 \pi^+}=\frac{4b_4^2}{15\pi f_{\pi}^2} \frac{M_{\Lambda_{b}^{0}}}{M_{\Sigma_{b}^{+}(1^2P_{\frac{3}{2}})}} p_{\pi}^5
\end{equation}
here, coupling constant $b_4=0.4 \times 10^{-3} MeV^-1$ \cite{GandhiDecay}
For $\Sigma_{b}^{+}(1^2P_{\frac{3}{2}}) $ to $\Sigma_{b}^{*+} \pi^0$ transition, decay rate is(coupling constant $b_5=\sqrt{2} \cdot b_1$ \cite{GandhiDecay}),
\begin{equation}
\label{eqn:24}
    \Gamma_{\Sigma_{b}^{+}(1^2P_{\frac{1}{2}}) \rightarrow \Sigma_{b}^{*+} \pi^0}=\frac{b_5^2}{10\pi f_{\pi}^2} \frac{M_{\Sigma_{b}^{*+}}}{M_{\Sigma_{b}^{+}(1^2P_{\frac{3}{2}})}} p_{\pi}^5
\end{equation}

The calculated decay rates for $P$-wave, $S$-wave and $D$-wave transitions of non-strange singly bottom baryons has been listed in Table \ref{Tab:Table 7} comparing with other references. In Some strong decay studies, $\Sigma_b^+$ and $\Sigma_b^-$ states are considered as single state (as considering $u$ and $d$ quark similar). Comparing to the decay width of them,strong decay width of reference \cite{Guo} is in range of $6.73-13.45$ Mev for the decay of $\Sigma_b$ to $\Lambda_b$ with single pion, while in present data, calculated decay width is 7-8 MeV. And for $\Sigma_b^*$ to $\Lambda_b$ with single pion, the strong decay width is $10.00-17.74$ Mev, while decay width for the same is 12.80 MeV in present data. For decay channel $\Sigma_b \rightarrow \Lambda_b \pi$, strong decay results in reference \cite{Zahra} and reference \cite{Majethiya2009} are 12.44 MeV and 4.63 MeV respectively. Present strong decay width is $7-8$ Mev for the same. The present strong decay width of the decay channel $\Sigma_b^* \rightarrow \Lambda_b \pi$ is $12.80$ MeV, comparing to the results of references \cite{Zahra} and \cite{Majethiya2009}, $22.85$ MeV and $8.74$ MeV consecutively.

\begin{table}
\centering
\caption{Single pion strong decay widths (in MeV)}
\label{Tab:Table 8}       
\begin{tabular}{lllll}
\hline\noalign{\smallskip}
    
Decay mode & Present & \cite{Limphirat} & \cite{Hwang} & \cite{Liang}\\
       \hline
           $P$-wave transitions\\
    \hline
        $\Sigma_{b}^{+}(1^2S_{\frac{1}{2}}) \rightarrow \Lambda_{b}^0 \pi^+$ & 7.08 & 7.1 & 4.35 & 5.08\\
        $\Sigma_{b}^{-}(1^2S_{\frac{1}{2}}) \rightarrow \Lambda_{b}^0 \pi^-$ & 8.04 & 9.0 & 5.77 & 6.11\\
        $\Sigma_{b}^{*+}(1^2S_{\frac{1}{2}}) \rightarrow \Lambda_{b}^0 \pi^+$ & 12.80 & 12.4 & 8.50 & 9.71\\
        $\Sigma_{b}^{*-}(1^2S_{\frac{1}{2}}) \rightarrow \Lambda_{b}^0 \pi^-$ & 12.80 & 14.6 & 10.44 & 10.99\\
    \hline
        $S$-wave transitions\\
        \hline
        $\Lambda_{b}^{0}(1^2P_{\frac{1}{2}}) \rightarrow \Sigma_{b}^+ \pi^-$ & 10.95 \\
        $\Lambda_{b}^{0}(1^2P_{\frac{1}{2}}) \rightarrow \Sigma_{b}^- \pi^+$ & 10.70 \\
        \hline
        $\Sigma_{b}^{+}(1^2P_{\frac{1}{2}}) \rightarrow \Lambda_{b}^0 \pi^+$ & 71.01 \\
        $\Sigma_{b}^{-}(1^2S_{\frac{1}{2}}) \rightarrow \Lambda_{b}^0 \pi^+$ & 69.72 \\
    \hline
        $D$-wave transitions\\
        \hline
         $\Lambda_{b}^{0}(1^2P_{\frac{3}{2}}) \rightarrow \Sigma_{b}^+ \pi^-$ & 0.0003 \\
        \hline
        $\Sigma_{b}^{+}(1^2P_{\frac{3}{2}}) \rightarrow \Lambda_{b}^0 \pi^+$ & 12.02 \\
        $\Sigma_{b}^{*+}(1^4P_{\frac{5}{2}}) \rightarrow \Lambda_{b}^0 \pi^+$ & 11.64 \\
        \hline
        $\Sigma_{b}^{+}(1^2P_{\frac{3}{2}}) \rightarrow \Sigma_{b}^{*+} \pi^0$ & 0.42 \\
    \hline
    \end{tabular}
\end{table}

\section{Summery}
In this work, masses of the radial and orbital states of non-strange singly bottom baryons has been calculated using Hypercentral Constituent Quark Model(hCQM) and screened potential has been employed with color coulomb potential to calculate the mass specta. The calculated mass spectra are listed in Table \ref{Tab:Table 2}-\ref{Tab:Table 5}. The predicted masses are compared to the masses obtained by other theoretical approaches. For, $\Lambda_b^0$, our results are in good agreement with experimental results and with the result obtained in reference \cite{Ebert2011} which is predicted by relativistic approach. The screening effect can be seen in Regge trajectories of $\Lambda_b^0$ and $\Sigma_b^{\pm}$ baryons, the linearly fitted lines are going to intersect for higher excited states as shown in Fig. \ref{Fig:Figure 2} and  Fig. \ref{Fig:Figure 3}.

The magnetic moment of ground states of non-strange singly bottom baryons, the transition magnetic moment(changing $J^P$ value $\frac{3}{2} \rightarrow \frac{1}{2}$ in same baryonic state) and transition decay width has been calculated and compared with other predictions. The single-pion strong decay width has been determined for $P$-wave, $S$-wave and $D$-wave transitions \cite{GandhiDecay}. The results for magnetic moment, transition decay width and strong decay width in Table \ref{Tab:Table 6}, Table \ref{Tab:Table 7} and  Table \ref{Tab:Table 8} respectively.

Hence, the mass spectra prediction using the simple basic constituent quark model is quite similar with prediction using relativistic approach and others. So we will extend this scheme to calculate the mass spectra and other properties of strange singly heavy bottom baryons. 

%
%


\begin{thebibliography}{}
%
%

\bibitem{PDG}
P.A. Zyla et al. (Particle Data Group), Prog. Theor. Exp. Phys. 2020, 083C01 (2020).

\bibitem{LHCb123}
R. Aaij et al. (LHCb Collaboration), Phys. Rev. Lett. \textbf{123},
152001 (2019).

\bibitem{CDF2012}
T. Aaltonen et al. (CDF Collaboration), Phys. Rev. D \textbf{85},
092011 (2012). 

\bibitem{LHCb122}
R. Aaij et al. (LHCb Collaboration), Phys. Rev. Lett. \textbf{122},
012001 (2019).
\bibitem{Moosavi}
S. Mohammad Moosavi Nejada, A. Armat, Eur. Phys. J. A \textbf{56}, 287 (2020).

\bibitem{Azizi102}
K. Azizi, Y. Sarac, H. Sundu, Phys. Rev. D \textbf{102}, 034007 (2020).

\bibitem{Liang2020}
W. Liang, Q. F. Lü, Eur. Phys. J. C  \textbf{80}, 690, (2020).

\bibitem{Yang}
H. M. Yang and H. X. Chen, Phys. Rev. D \textbf{102}, 079901(E) (2020).

\bibitem{Wang2021}
Z. G. Wang, H. J. wang, Chin. Phy.s C \textbf{45}, 013109 (2021).

\bibitem{Azizi101}
K. Azizi, Y. Sarac, H. Sundu, Phys. Rev. D \textbf{101}, 074026 (2020).

\bibitem{XChen}
H. X. Chen, W. Chen, Q. Mao, A. Hosaka et. al., Phys. Rev. D \textbf{91}, 054034 (2015).

\bibitem{Thakkar2017}
K. Thakkar, Z. Shah, A.K. Rai and P.C. Vinodkumar, Nucl. Phys. A \textbf{965}, 57 (2017).

\bibitem{Zahra2011}
N. Salehi, A. A. Rajabi, Z. Ghalenovi, Acta PhysicaPolonoca B, \textbf{42} (2011).

\bibitem{Ebert2011}
D. Ebert, R. N. Faustov and V. O. Galkin, Phys. Rev. D \textbf{84}, 014025 (2011).

\bibitem{BChen}
 B. Chen, K.-W. Wei, and A. Zhang, Eur. Phys. J. A 51, 82
(2015).
 
\bibitem{CapsticIsgur}
 S. Capstick and N. Isgur, Phys. Rev. D 34, 2809 (1986).

\bibitem{Jia}
D. Jia, W.N. Liu and A. Hosaka, Phys. Rev. D \textbf{101}, 034016 (2020).

\bibitem{Kim}
Y. Kim, E. Hiyama et al., Phys.Rev.D \textbf{102}, 014004 (2020).

\bibitem{ShahEPJA2016}
Z. Shah, Thakkar K., Rai A. K., VinodkumarP. C., Eur. Phys. J A \textbf{52}, 313 (2016).

\bibitem{Yamaguchi2015}
Y. Yamaguchi, S. Ohkoda, A. Hosaka, T. Hyodo and S. Yasui, Phys. Rev. D \textbf{91}, 034034 (2015).

\bibitem{Zhong}
X. H. Zhong, Q. Zhao, Phys. Rev. D \textbf{77}, 074008 (2008).

\bibitem{Namekawa}
Y. Namekawa, S. Aoki, K. I. Ishikawa, N. Ishizuka et. al., Phys. Rev. D \textbf{87}, 094512 (2013).

\bibitem{Brown}
Z. S. Brown, W. Detmold, S. Meinel, K. Orginos, Phys. Rev. D \textbf{90}, 074507, (2014).

\bibitem{Alexandrou}
C. Alexandrou, V. Drach, K. Jansen, C. Kallidonis, G. Koutsou, Phys. Rev. D \textbf{90}, 074501 (2014),

\bibitem{Padmanath}
M. Padmanath, N. Mathur, arXiv:1508.07168v1 [hep-lat] (2015).

\bibitem{Can}
K. U. Can, G. Erkol, M. Oka, T. T. Takahashi, Phys. Rev. D \textbf{92}, 114515 (2015).

\bibitem{Mannel}
T. Mannel, W. Roberts, Z. Ryzak, Nucl. Phys. B \textbf{355}, 38-53 (1991).

\bibitem{Giannini}
 M.M. Giannini and E. Santopinto, Chin. J. Phys. \textbf{53}, 020301 (2015).

\bibitem{Bijkar2000}
R.  Bijkar, F. Iachello, A. Leviatan, Ann. Phys. \textbf{284}, 89 (2000).

\bibitem{Wang2019}
J. Z. Wang, D. Y. Chen,X. Liu, T. Matsuki, Phys. Rev. D \textbf{99}, 114003 (2019).

\bibitem{Li2009}
 B. Q. Li, K. T. Chao , Phys. Rev. D \textbf{79}, 094004 (2009).

\bibitem{Voloshin2008}
M. B. Voloshin , Prog. Part. Nucl. Phys. \textbf{61}, 455 (2008).

\bibitem{Bijkar1994}
 R. Bijkar, F. Iachello, A. Laviatan, Ann. Phys. (N. Y.) \textbf{236}, 69 (1994).

\bibitem{Bijkar1998}
R. Bijkar, F. Iachello, E. Santopinto, J. Phys. A \textbf{31}, 9041 (1998).

\bibitem{Lucha1999}
W. Lucha and F. Schoberls, Int. J. Mod. Phys. C \textbf{10}, 607 (1999).

\bibitem{Gandhinon-strange}
K. Gandhi ,Z. Shah, A. K. Rai, Int. J Theor. Phys. \textbf{59}, 1129–1156 (2020).

\bibitem{CDF2007}
T. Aaltonen et al. (CDF Collaboration), Phys. Rev. Lett. \textbf{99},
202001 (2007).



\bibitem{DELPHI}
P. Abreu et al. (DELPHI Collaboration), Phys. Lett.
B374, 351 (1996).

\bibitem{ALEPH}
D. Buskulic et al. (ALEPH Collaboration), Phys. Lett.
B380, 442 (1996).


\bibitem{GandhiDecay}
K. gandhi, Z. Shah, A. K. Rai, Eur. Phys. J. Plus \textbf{133}, 512 (2018).

\bibitem{Majethiya2016}
A. Majethiya, K. Thakkar, P.C. Vinodkumar, Chin. J. of Phys., \textbf{54}, 495-502 (2016).

\bibitem{Aliev}
T. M. Aliev, K. Azizi and  A. Ozpinecix, Phys. Rev. D, \textbf{79}, 056005 (2009).

\bibitem{Pirjol}
D. Pirjol, T. M. Yan, Phys.Rev. D \textbf{56}, 5483 (1997).

\bibitem{Limphirat}
A. Limphirat et al., arXiv:0710.3942[hep-ph].

\bibitem{Hwang}
C.W. Hwang, Eur. Phys. J. C, \textbf{50}, 793 (2007).

\bibitem{Liang}
W. Liang, Q. F. Lü, and X. H. Zhong,  Phys. Rev. D, \textbf{100}, 054013 (2019).

\bibitem{Guo}
X. H. Guo , et al. , Phys. Rev. D, \textbf{77} (2008) 036003 . 

\bibitem{Zahra}
Z. Ghalenovia and M. M. Sorkhi, Eur. Phys. J. Plus \textbf{133}: 301 (2018).

\bibitem{Majethiya2009}
A. Majethiya, B. Patel and P. C. Vinodkumar, Eur. Phys. J. A \textbf{42}, 213-218 (2009).

\bibitem{Roberts}
 W. Roberts and M. Pervin, Int. J. Mod. Phys. A \textbf{23}, 2817-2860 (2008).

\bibitem{WHLiang}
W. H. Liang, C.W. Xiao, and E. Oset, Phys. Rev. D \textbf{89}, 054023 (2014).

\bibitem{Bhavsar}
N. Bhavsar, M. Shah et.al., Proceedings of the DAE Symp. on Nucl. Phys. \textbf{63}, (2018).









\end{thebibliography}


\end{document}